\documentclass[sigconf]{acmart}

\AtBeginDocument{%
  \providecommand\BibTeX{{%
    \normalfont B\kern-0.5em{\scshape i\kern-0.25em b}\kern-0.8em\TeX}}}

\copyrightyear{2022}
\acmYear{2022}
\setcopyright{acmcopyright}
\acmConference[ASSETS '22]{The 24th International ACM
SIGACCESS Conference on Computers and Accessibility}{October 23--26,
2022}{Athens, Greece}
\acmBooktitle{The 24th International ACM SIGACCESS Conference on Computers and Accessibility (ASSETS '22), October 23--26, 2022, Athens, Greece}
\acmPrice{15.00}
\acmDOI{10.1145/3517428.3544829}
\acmISBN{978-1-4503-9258-7/22/10}

\graphicspath{{figures/}}

\usepackage{csquotes}
\renewcommand{\mkbegdispquote}[2]{\itshape}
\usepackage{enumitem}

\begin{document}

\title[Avatar Diversity and the Self-presentation of People with Disabilities in Social VR]{``It's Just Part of Me:'' Understanding Avatar Diversity and Self-presentation of People with Disabilities in Social Virtual Reality}

\author{Kexin Zhang}
\affiliation{
    \institution{University of Wisconsin-Madison}
    \city{Madison}
    \state{Wisconsin}
    \country{USA}
}
\email{kzhang284@wisc.edu}

\author{Elmira Deldari}
\affiliation{
    \institution{Univ. of Maryland, Baltimore County}
    \city{Baltimore County}
    \state{Maryland}
    \country{USA}
}
\email{edeldar1@umbc.edu}

\author{Zhicong Lu}
\affiliation{
    \institution{City University of Hong Kong}
    \city{Kowloon}
    \state{Hong Kong}
    \country{China}
}
\email{zhicong.lu@cityu.edu.hk}

\author{Yaxing Yao}
\affiliation{
    \institution{Univ. of Maryland, Baltimore County}
    \city{Baltimore County}
    \state{Maryland}
    \country{USA}
}
\email{yaxingyao@umbc.edu}

\author{Yuhang Zhao}
\affiliation{
    \institution{University of Wisconsin-Madison}
    \city{Madison}
    \state{Wisconsin}
    \country{USA}
}
\email{yuhang.zhao@cs.wisc.edu}








\renewcommand{\shortauthors}{Zhang et al.}

\begin{abstract}
In social Virtual Reality (VR), users are embodied in avatars and interact with other users in a face-to-face manner using avatars as the medium. With the advent of social VR, people with disabilities (PWD) have shown an increasing presence on this new social media. With their unique disability identity, it is not clear how PWD perceive their avatars and whether and how they prefer to disclose their disability when presenting themselves in social VR. We fill this gap by exploring PWD's avatar perception and disability disclosure preferences in social VR. Our study involved two steps. We first conducted a systematic review of fifteen popular social VR applications to evaluate their avatar diversity and accessibility support. We then conducted an in-depth interview study with 19 participants who had different disabilities to understand their avatar experiences. Our research revealed a number of disability disclosure preferences and strategies adopted by PWD (e.g., reflect selective disabilities, present a capable self). We also identified several challenges faced by PWD during their avatar customization process. We discuss the design implications to promote avatar accessibility and diversity for future social VR platforms.
\end{abstract}

\ccsdesc[500]{Human-centered computing~Virtual reality}
\ccsdesc[500]{Human-centered computing~Accessibility technologies}

\keywords{Social VR, avatar, self-perception, disability disclosure, visual impairments, d/Deaf and heard of hearing}

\maketitle

\section{Introduction}

With the rising popularity of virtual reality (VR), social VR is becoming the mainstream of the online social eco-system. Social VR refers to VR platforms where people communicate and socialize in the form of avatars~\cite{freeman2020my}. One iconic characteristic of social VR is the embodied ``face-to-face'' interaction from a first-person perspective through avatars.
By customizing their avatars, users can craft and maintain any characters they prefer in a virtual world.

Due to the deep embodiment of a user with their avatars, to some degree, a customized avatar can be considered as a proxy of the user themselves~\cite{manninen2007value}. A myriad of research has explored how people use avatars to shape their self-images in virtual social spaces, including both PC-based virtual worlds \cite{kafai2007your} and the more immersive social VR \cite{freeman2021body, freeman2020my}. Research has found that while many people designed their avatars to reflect their physical appearance in the real world \cite{martey2011performing}, some people leveraged the opportunity to experiment with different aspects of their personalities \cite{fong2015does, ducheneaut2009body} and even explored completely different identities \cite{waggoner2009my}. However, most prior works focused on the self-presentation preferences of people without disabilities. People with disabilities (PWD), given their unique disability identity, may have different perceptions and preferences when constructing self-images via avatars in social VR.

Research on PWD's self-presentation in social VR is in its infancy. On traditional social media platforms, PWD have a strong presence and use various strategies to construct their online images as others do \cite{wu2014visually, shpigelman2014facebook}. Unlike people with typical abilities, disability disclosure is a unique perspective that PWD need to consider during self-presentation. Many PWD are cautious about disclosing their disability-related vulnerabilities, and some intentionally hide their disability to avoid potential risks, such as loss of job opportunities ~\cite{zhao2017effect} and cyber-bullying \cite{kowalski2016cyberbullying}. In contrast, some proactively disclose their disabilities, especially on dating platforms, to build trust and filter potential partners \cite{porter2017filtered}. Unlike the 2D text- or image-based social media, social VR provides a more embodied and immersive experience via avatars, which can potentially change PWD's self-presentation preferences and their willingness to disclose their disabilities. To promote safe and inclusive social VR experiences for PWD, the accessibility and VR communities are in need of a thorough understanding of PWD's avatar perceptions and disability disclosure preferences on the emerging social VR platforms.

Our research aims to fill the gap by investigating how PWD design and use their avatars for disability disclosure and self-presentation in social VR. Given that social VR is an emerging but premature medium that lacks sufficient accessibility support, we also explore the status quo of avatar diversity and accessibility on commercial social VR platforms to reveal the barriers faced by PWD in the avatar customization process. Specifically, we aim to answer the following research questions:

\begin{itemize}
    \item \textbf{RQ1:} Whether and how is avatar diversity supported on the mainstream commercial social VR platforms?
    \item \textbf{RQ2:} Whether and how do PWD disclose their disabilities when presenting themselves via avatars?
    \item \textbf{RQ3:} What challenges do PWD face during the avatar design and creation process?
\end{itemize}

To answer these research questions, we first systematically reviewed 15 commercial social VR applications to understand what avatar features are supported for disability representation. We then conducted in-depth semi-structured interviews with eight visually impaired people, nine d/Deaf and hard of hearing people, and two people with multiple disabilities
to explore their avatar perceptions and customization experiences in the social VR context. 
Our findings revealed a spectrum of disability disclosure strategies adopted by PWD for self-presentation, from accurately disclosing one's disability, to selectively presenting a particular aspect of the disability, presenting the changes in one's ability, to hiding one's disability to construct a different self. Our study highlighted the lack of avatar diversity for disability representation and emphasized PWD's needs to flexibly control their disability disclosure in different social contexts. 
We also identified the avatar accessibility barriers faced by PWD and suggested design implications.

This paper makes three contributions. First, to the best of our knowledge, this is the first research that investigates social VR avatar diversity and self-presentation from the lens of disability. Second, our research presents rich and in-depth data to uncover PWD's different disability disclosure strategies and preferences. Third, design implications are derived to inspire a more accessible and inclusive avatar experience for PWD in social VR.  

\section{Related Work}
\subsection{Self-presentation in Virtual Worlds}
Self-presentation, also known as impression management, refers to the ways that a person ``conveys an impression to others which is in his interests to convey'' \cite{goffman2021presentation}. Via self-presentation, people can selectively craft, present, and maintain specific facets of their identities based on different audiences and social settings \cite{abrams2004metatheory, farnham2011faceted, kairam2012talking, Neustaedter2009}. A myriad of research has explored how people manage their identities on social media (e.g., Facebook, Twitter), revealing the complexity of self-presentation from various aspects, such as beautified real self on anonymous social platforms \cite{zhao2008identity, gibbs2006self, yurchisin2005exploration}, multiple online identities for different audience groups \cite{dimicco2007identity}, and the stereotypical representation of genders \cite{bailey2013negotiating}.

With the advances in computer graphic technology and hardware support, social media expands from 2D to 3D avatar-based systems \cite{freeman2021body}. Instead of using text and images to construct self-image, people can directly create and personalize an avatar as their virtual representation and manage the impression from others via the avatar. Many researchers explored users' digital representations via avatars in PC-based virtual worlds, including both the game worlds (e.g., World of Warcraft \cite{WoW}) and the social worlds (e.g., Second Life \cite{SecondLife}). In the virtual game worlds where a limited number of pre-defined avatars were provided, many users tended to select avatars that stood out, followed a trend \cite{ducheneaut2009body}, or served the role they planned to play in the game \cite{gee2003video}. In contrast, social worlds gave users more flexibility to design their own avatars, for example, by adjusting and customizing different body features (e.g., hair, eyes, skin color). Without restrictions from the game theme, many users chose to create avatars that resembled themselves in the real world \cite{ducheneaut2009body, serapis2008coming, wallace2009impact}. For example,
Ducheneaut et al. \cite{ducheneaut2009body} explored users' avatar customization across three virtual worlds (i.e., Maple Story, World of Warcraft, and Second Life), and found that more users preferred reproducing some of their physical characteristics in the social virtual worlds than in the game worlds. Koles et al. \cite{koles2012portrayed} also showed that the majority of users in Second Life tended to use their physical selves as the starting point to create their virtual representation.

With the high flexibility in avatar customization, people also have the freedom to create avatars that are different from their physical appearances, thus better shaping their online identities. As a result, some people used avatars to experiment with different aspects of their personalities \cite{turkle1999life, turkle2005second, kafai2010your}. For example, Kafai et al. \cite{kafai2010your} surveyed 438 tween players in a virtual social community named \textit{Whyville} and found that most tweens did not construct their avatar appearances based on their real selves. Instead, they designed avatars to achieve recognition from others or to reveal specific aspects of their ``real'' selves, including those they desired but could not present in real life. Via avatar creation, users can even create and explore a completely different identity, such as an avatar with a different gender \cite{hussain2008gender} or a different race \cite{nakamura1995race}, which may help them discover their true selves and increase their self-esteem \cite{bessiere2007ideal}. 

Most prior work on avatar perception focused on the majority group. Limited attention has been paid to the self-presentation and identity disclosure of the under-representative user groups, especially people with disabilities.   

\subsection{Self-disclosure by People with Disabilities in Virtual Worlds}
Different from self-presentation where people can construct and present any images that they prefer to the audience, self-disclosure refers to the act of revealing any messages about oneself to others \cite{gibbs2006self, greene2006self, cozby1973effects}. While self-disclosure is involved in self-presentation strategies to help build closer relationships \cite{SCHLOSSER20201}, it also poses potential risks of exposing one's vulnerabilities, especially to the unknown or anonymous audience online \cite{ma2016self}. Prior research has explored the online self-disclosure experiences of various under-representative groups, such as racial minority \cite{lee2014does, lee2011whose, higgin2009blackless} and gender minority \cite{justin2021trans, cassell1998chess, eden2010gender, morgan2020role}. In this section, however, we focus on the work for people with disabilities (PWD). 

Disability disclosure has always been an important topic for PWD even before social media platforms emerged \cite{huvelle1984tell, pearson2003tell, von2014perspectives, riddell2014disabled}.
Prior research focused primarily on the physical working context, showing that disability disclosure could potentially assure appropriate workplace accommodations and increase workplace diversity and inclusiveness for PWD \cite{von2014perspectives}. However, it may also lead to negative employment consequences for PWD, such as lowered supervisor expectations, isolation from colleagues, and increased possibility of termination \cite{huvelle1984tell, pearson2003tell}. Because of these, in working or daily living contexts, disability disclosure is highly contextualized, and influenced by many factors such as employers, managers, and workplace climate \cite{von2014perspectives, pearson2003tell}. 

The advent of online social platforms (e.g., social media, social virtual worlds) affords new forms of social interactions and activities, along with new social relationships. These online platforms bring PWD more opportunities to interact with others and get access to resources and communities \cite{stendal2012ICTreview, stewart2010}. For example, Boellstorff \cite{Boellstorff2019} built ``Ethnographia'', a virtual island in Second Life, to investigate how digital spaces influenced PWD's experiences and found that building virtual worlds enhanced PWD's sense of ability and helped them reconstruct identities.
With these online social platforms, PWD were able to manage their disability disclosure, thus achieving a ``levelling ground'' where they could be treated on their merits as a person without being shadowed by their disabilities \cite{bowker2002disability}. However, disability disclosure in online communities can also bring risks to PWD \cite{stendal2012ICTreview, ringland2019}. For example, Ringland \cite{ringland2019} conducted a 200-hour observation in Autcraft, an online community for children with autism, and found that the autistic identity in virtual worlds could be both a source of empowerment and a source of harassment and violence.

Researchers have explored PWD's self-disclosure strategies and preferences in different online communities \cite{porter2017filtered, gerling2016design, ringland2019,edwards2021image, Boellstorff2019}. 
For example, Porter et al. surveyed 91 adults with and without disabilities to understand the needs for disability disclosure on online dating platforms. They found a higher expectation of the disclosure of visible disabilities than invisible disabilities \cite{porter2017filtered}.
By designing a movement-based virtual game for young people using wheelchair and exploring their avatar preferences, Gerling et al. found that while 6 out of 8 participants depicted wheelchairs as indispensable parts of their self-images, only two participants were willing to use avatars with mobility disabilities in the game \cite{gerling2016design}. Recently, Davis and Stanovsek conducted a 3-year ethnography study to explore how PWD customize their avatars in Second Life and found that many participants used non-human avatars to free themselves from their visible disabilities \cite{davis2021machine}. 

Compared to the conventional social media and virtual worlds, the emerging social VR brings unique embodied experiences, which may potentially affect PWD's willingness of disability disclosure. However, little work has investigated the influence of social VR on PWD's self-presentation and disability disclosure preferences.  

\subsection{Avatars Design and Self-presentation in Embodied Social VR}
In recent years, social VR has gained increasing attention. Unlike the PC-based virtual worlds where users see avatars on a 2D screen and can only control the avatars via a keyboard and a mouse, social VR incorporates full-body tracking, providing a more embodied first-person avatar experience as well as richer and more immersive social interactions \cite{freeman2021body}. 
As a result, social VR provides new affordances for avatar design and interaction \cite{hepperle2021aspects, menon2021role, kolesnichenko2019understanding, waltemate2018impact}. For instance, Kolesnichenko et al. \cite{kolesnichenko2019understanding} interviewed industry experts who work for different commercial social VR platforms and uncovered the current avatar design practices from different aspects, such as locomotion, avatar aesthetics, and avatar's relation to one's virtual identity. Menon \cite{menon2021role} also researched the relationship between avatar customization and embodiment in a virtual job interview context, highlighting the needs for basic avatar customization.

Some research specifically focused on avatar-based self-presentation in social VR and how it affected users' behaviors and interactions \cite{freeman2020my, freeman2021body, maloney2020anonymity}. For example, Freeman et al. \cite{freeman2021body} interviewed 30 people about their avatar perception and social interaction experiences in social VR. They found that most people considered the avatars to be themselves and strived to make their avatars similar to their physical appearances. This research also revealed how avatar genders and race affected people's experience. For example, female avatars received better first impression and nicer treatment but also led to more harassment, and non-white avatars could bring certain social stigma. Moreover, Maloney and Freeman \cite{maloney2020anonymity} studied how and why people disclose their information in social VR. They argued that creating avatars that look like one's physical self can disclose important personal information such as gender, race, and appearance. While this helped build close connection with others, it can potentially lead to privacy risks. 

With more realistic avatars and more embodied interactions, social VR can also affect PWD's self-presentation and disability disclosure preferences.  Some researchers have noticed the importance of this research direction. Boellstorff \cite{Boellstorff2019} emphasized the importance of distinguishing ``virtual world'' and ``virtual reality'' for understanding PWD's avatar experience.
Mott et al. \cite{Mott2019oppo} also stressed the needs for supporting more diverse avatar representations for PWD in social VR. However, to our knowledge, no research has thoroughly explored PWD's avatar design experiences in the embodied social VR. 
Our research contributes to this line of research by developing
a deep understanding of the current avatar diversity practices, as well as PWD's self-presentation challenges and strategies in social VR.

\section{Study I: Application Review: Avatar Diversity on Commercial Social VR Platforms}
The goal of this study is to build a comprehensive understanding of the avatar diversity and accessibility practices on current commercial social VR platforms (RQ1). Specifically, we aim to uncover 1) the \textit{general avatar design practices}: what types of avatars are supported and how users can select or customize an avatar on different social VR platforms; 2) \textit{avatar diversity support}: what avatar features are provided to enable disability disclosure for PWD; and 3) \textit{avatar accessibility}: what accessibility features are provided for PWD to perceive, use, and customize avatars. 

\subsection{Method}
We conducted a systematic review of fifteen popular commercial social VR applications.
To determine which social VR applications to review, we conducted an exhaustive search on three mainstream VR application stores, including Oculus, Viveport, and Steam. Our search focused on applications available in the United States in November and December 2021. We first searched the keyword ``social'' in these stores and identified a total of 133 VR applications: 47 from Oculus, 54 from Viveport, and 50 from Steam (18 apps were available across multiple stores). 
To further narrow down the scope, we filtered the applications by checking the application descriptions and only focused on the applications whose descriptions clearly indicated a social or collaborative nature.
Finally, we excluded VR applications that showed an intense gaming nature (e.g., sports, shooting games), which can distract users from socializing, such as EchoVR, StarTrek: Bridge Crew, and SpaceteamVR. We also consulted other related work and online articles about popular social VR applications to ensure not missing any mainstream social VR platforms \cite{liu2021socialvr, 2022bestsocial, 2021bestsocial}. This process resulted in fifteen social VR applications with strong social nature, including Rec Room, VRChat, Horizon Worlds, vTime XR, AltspaceVR, Bigscreen, Alcove, Half + Half, Horizon Venues, Villa, Arthur, ENGAGE, Multiverse, PokestarVR, and Spatial (details can be found in Table \ref{tab:socialVRapps} in Appendix).

Two researchers on the team reviewed all applications independently. We adopted a depth-first traverse strategy, clicking all available buttons and menu items in the avatar customization process. During the review, researchers video-recorded and took notes of all avatar options and the interaction process. The two researchers then discussed their results to ensure the reliability of the review. Following the same strategy, we reviewed the general setting in each social VR application to examine what accessibility features were supported. Any accessibility features that could influence or be applied to the avatar selection and customization process were documented. All applications were reviewed with Oculus Quest 2. 

\subsection{Review Results}
\subsubsection{\textbf{Avatar Creation Methods.}}
We identified four ways to create and customize avatars in social VR: (1) \textit{Full Avatar Selection}: selecting from a set of pre-determined avatars provided by the system, (2) \textit{Avatar Feature Customization}: customizing different components of a human avatar (e.g., eyes, hair styles), (3) \textit{Photo-based Avatar Generation:} uploading a photo to the system and generating a corresponding avatar automatically, and (4) \textit{Third-party Avatar 
Import}: designing an avatar using a third-party platform and uploading it to the social VR application.

Most social VR platforms (12 out of 15) supported Avatar Feature Customization, allowing users to customize the avatar's physical features (e.g., skin tone, body shape, eyes), clothing (e.g., outfits, accessories), or both. Three platforms (VRChat, Half+Half, and Multiverse) employed the Full Avatar Selection model. Notably, VRChat offered a set of 80 pre-defined avatars for users to select and supported various avatar types, including both humanoid avatars and non-humanoid avatars, such as cartoons, robotic figures, animals, and objects. VRChat was also the only platform that enabled users to design and import their own avatars from third-party platforms. Additionally, three platforms (Villa, Arthur, and Spatial) employed Photo-based Avatar Generation.
Table \ref{tab:socialVRapps} in Appendix listed the avatar customization methods supported by different platforms.

Beyond individual applications, the Oculus system provided its own avatar system (i.e., Meta Avatars) that employed the Avatar Feature Customization model. When first logged in to an Oculus device, users were automatically directed to Horizon Home\footnote{A virtual home in Oculus Quest 2, which functions like the homepage of 2D platforms and enables certain social activities \cite{horizon}.} to customize their avatars for the whole VR system. Three platforms--Horizon Worlds, Horizon Venues, Alcove--directly adopted the Meta Avatars, so that a user could continue using the same avatar from the Oculus system. 

Our review of the avatar customization process confirmed the results from Kolesnichenko et al.'s work \cite{kolesnichenko2019understanding} and further extended the scope to more widely-used social VR applications.

\subsubsection{\textbf{Avatar Realism.}}
Most social VR platforms provided only humanoid avatars. We investigated the avatar realism by examining the completeness of the humanoid avatars. We found that only three platforms--VRChat, vTime XR, and ENGAGE--provided full body avatars, including head, neck, torso, arms, hands, legs, and feet. All other 12 platforms focused on rendering the upper body of the avatars, presenting at least the avatars' head and hands. Specifically, five platforms did not show the avatars' arms but only presented flowing hands; avatars in Rec Room and Bigscreen did not have necks; and Villa's avatars did not even have a torso. In terms of avatar rendering details, except for Rec Room, all social VR platforms we reviewed rendered avatar fingers and tracked the users' finger movement via the VR controllers. Table \ref{tab:socialVRapps} in Appendix provides details of the avatar realism on all platforms.

\subsubsection{\textbf{Disability Representation in Avatars}}
\begin{figure} [h]
    \centering
    \includegraphics[width=\linewidth]{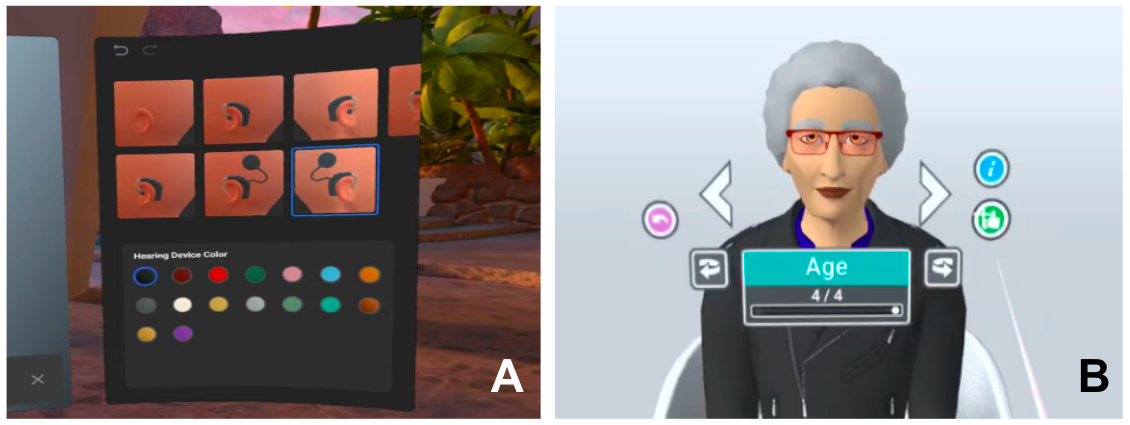}
    \caption{Avatar Diversity: A. Hearing devices in Meta Avatars; B. Sagging facial skin in vTime XR Avatar.} 
    \Description{Fig. 1A.  showed Meta Avatars’ seven ways in wearing hearing aids and cochlear implants. There are seven squares, and each square shows avatar’s one-side ear of either wearing a hearing device or not. 16 colors are provided at the bottom of figure for users to customize the color of hearing devices. Fig. 1B. illustrated a female avatar in vTime XR who looks over age of 50. She is sitting on a chair and facing forward. Her hair and eyebrows are in color of white, and she has droopy eyelids and sunk cheeks.}
    \label{fig:avatar_diversity}
    \vspace{-3ex}

\end{figure}

Our review uncovered the limited avatar diversity support for PWD. We found that most commercial social VR platforms did not offer any disability-related avatar features. Meta Avatars was the only avatar system that provided hearing device features for people who are d/Deaf or hard of hearing. Two types of hearing devices were supported: cochlear implants and hearing aids. A user can put the hearing device on the avatar via three options: left ear only, right ear only, and both ears (Fig \ref{fig:avatar_diversity}A). This feature was provided under the category of ``Clothing.'' However, no other assistive devices or disabilities were supported. Additionally, we found some ``near disability-related'' features that may be used to indicate a disability. Specifically, Bigescreen provided an eye patch feature under ``Glasses'' category, which users can add onto their avatars to present their visual impairments or eye injuries. However, we acknowledge that this feature might be designed as an eye decoration (e.g., to imitate a pirate) instead of a disability feature.  

Besides disabilities, some applications allowed users to customize wrinkles and face lines on avatars to reflect age and represent older adults. For example, Meta Avatars provided five levels of face lines with different depth and number of wrinkles on avatar's face to demonstrate different ages. vTime XR offered four sagging levels for avatars' facial skin to present age (Fig \ref{fig:avatar_diversity}B).

\subsubsection{\textbf{Accessibility Features for Avatar Customization}}
The accessibility features on social VR platforms were limited. While some platforms supported standard accessibility features that were commonly used on computer and smartphone interfaces (e.g., magnification, color correction), no features were designed specifically for avatar customization. Specially, AltspaceVR allowed users to adjust the scale of the avatar interface. The Oculus system also provided visual augmentations (i.e., color correction, text size adjustment) that can be applied to any applications on Oculus devices. 
Moreover, most social VR platforms allowed adjustment to the audio and haptic feedback, which could be helpful for PWD. 

By reviewing mainstream social VR platforms, we developed a comprehensive understanding of the current practices of the avatar design, diversity, and accessibility. The results of this study inspired and served as a solid grounding for our following interview study (Study II). 
First, the review identified existing disability features for avatars and the social VR platforms that supported such features, which enabled us to better understand participants' social VR experiences and disability disclosure choices in an in-depth interview. Second, the limited disability features (i.e., hearing aids, eye patches) identified in this study helped us narrow down the focus to participants who were visually impaired and deaf or hard of hearing, since they were the only disability groups that had avatar representations in social VR.

\section{Study II: Interviews: Self-Presentation of People with Disabilities via Avatars} \label{Study2}
In this study, we investigated how PWD design and craft their avatars to disclose their disabilities and present themselves in social VR (RQ2) as well as their challenges and needs during the avatar creation and customization process (RQ3). 

\subsection{Method}
We conducted semi-structured interviews with 19 PWD via Zoom in February and March 2022. The study was approved by the Institutional Review Board (IRB).

\subsubsection{Participants}
Our recruitment focused on two types of sensory disabilities, (1) d/Deaf or Hard of Hearings (DHH) people and (2) visually impaired (VI) people, since they are common and visible disabilities (or disabilities that can be indicated by visible assistive technologies). Both disabilities can be easily noticed in the real world and thus can be reflected via avatar design. For example, blind people may use a white cane, and d/Deaf people may wear hearing aids or may sign. We recruited 19 participants (11 female, 7 male, and 1 transgender). Their ages ranged from 20 to 58 ($mean=33, SD=11.57$). Among the participants, eight had visual impairments (V-P1 to V-P8), nine were DHH (H-P1 to H-P9), and two had multiple disabilities (M-P1, M-P2). M-P1 was both blind and DHH, while M-P2 was blind and had a prosthetic leg. Table \ref{tab:my_label} 
shows participants' detailed information. 

We spread our recruitment information via (1) the mailing lists of non-profit disability organizations, such as the National Federation of the Blind and National Association of the Deaf, 
and (2) mainstream social media platforms, such as Facebook groups, Twitter, and Instagram. Interested participants could fill in a screening survey with their age, disability conditions, and general avatar experiences. Participants were eligible if they were over 18 and had visual impairments or were d/Deaf or hard of hearing. We limited the recruitment to individuals who spoke English.
If selected to participate, participants were asked to sign a consent form prior to the interview. Upon completion, participants received \$15 as compensation. 

\subsubsection{Interview protocol}
The interview included three phases. The first phase focused on participants' background information, including demographics (age and gender), self-reported disability, and experiences with VR and social VR. 

The second phase focused on participants' self-presentation and disability disclosure via avatars. We first asked participants to send photos of their avatars or screen share their avatars with us to demonstrate their avatar design. We asked about how they designed their avatars, why they customized their avatars in this way, and whether their avatars involved any features to disclose their disabilities. If the participants disclosed their disabilities via their avatars, we further asked how they disclosed their disabilities, why they wanted to do so, and their experience in social VR after they disclosed disabilities via avatars. If participants' avatars did not indicate their disabilities, we asked about their willingness of disclosing their disabilities via avatars and the reason. 
Additionally, we asked participants whether and how they had disclosed their disabilities through any other means or on any other online social platforms and the rationales. 
Participants also discussed whether and how they wanted the social VR platforms to support disability representations via avatars. 

The last phase of the interview focused on avatar creation accessibility. 
We asked about participants' experiences, difficulties, and strategies when creating and customizing avatars and the types of assistance they needed to complete the avatar creation process. We also asked about their suggestions for a more accessible avatar experience.

\subsubsection{Data recording and analysis }
Upon participants' consent, all 19 interviews were recorded and auto-transcribed by Zoom. Participants had the option to turn off their cameras during the interview, although it was not required. Two researchers manually cleaned all transcripts by checking the recorded interviews.  

We conducted thematic analysis~\cite{clarke2015thematic, braun2006using} to identify repetitive patterns and themes in the interviews. 
First, we manually organized all qualitative data in an Excel sheet and selected five representative transcripts (2 DHH participants, 2 VI participants, and 1 participant with multiple disabilities) as samples. Three researchers coded all samples independently at the sentence level with open coding. Then, they discussed and reconciled their codes to resolve any differences, and developed an initial codebook upon agreement. Next, two researchers divided the rest of the transcripts based on the participants' disabilities. Specifically, one researcher analyzed all VI participants' data (including the two participants with multiple disabilities) and the other researcher analyzed all DHH participants' data. During this process, the two researchers regularly checked each other's codes and discussed as needed to ensure consistency. New codes were added to the codebook based on the agreement between the two researchers. In the meantime, the third research oversaw all these activities to ensure a high-level agreement. The final codebook contained over 120 codes. We categorized all the codes into high-level themes using affinity diagram and achieved twenty themes.

\begin{table*}[h!] 
    \footnotesize 
    \centering
    \caption{Participants' demographics and social VR experiences. * indicates avatar platforms demonstrated by our participants.}
    \begin{tabular}{p{0.6cm}p{0.8cm}p{3cm}p{2.2cm}p{1.8cm}p{4.6cm}}
    \toprule
      \textbf{ID}&  \textbf{Age/ \newline Sex}&  \textbf{Self-reported Disability}&  \textbf{Assistive Tech Used}&  \textbf{Social VR \newline Experience}&   \textbf{Social VR Platforms Used}\\
     \midrule 
    
   \textbf{H-P1}&   52/F&   Profound deaf since birth &  Cochlear implant&  12 months&  VRChat, AltspaceVR, Spatial, Mozilla Hubs, Rec Room, Meta Avatars*\\ 
    \midrule
    
    \textbf{H-P2}&   20/F&   Severe deaf in left ear&  Hearing aid&  Multiple times&  A social VR game, Snapchat*\\ 
    \midrule

     \textbf{H-P3}&   20/F&   25\% hearing loss in right ear&  Hearing aid&  Tried once&   VR Exhibition at a museum, Snapchat*, Meta Avatars*, Sims*, Memoji*\\ 
    \midrule
    
      \textbf{H-P4}&    23/Trans&    Moderate hearing loss &   Hearing aids&   24 months&    VRChat*\\ 
    \midrule
    
      \textbf{H-P5}&    29/F&    Moderate hearing loss &   Hearing aids&   Multiple times&   VRChat, Rec Room, Snapchat*, Meta Avatars*, Roblox*\\ 
    \midrule
    
      \textbf{H-P6}&    25/M&    Profound deaf &   N/A&   12 months&   VRChat*, SteamVR\\ 
    \midrule
    
      \textbf{H-P7}&    30/M&    Profound deaf &   Cochlear implant&   12 months&   VRChat, Rec Room*, Lost Horizon, Meta Avatars*\\ 
    \midrule
    
      \textbf{H-P8}&    25/M&    Deaf &   Hearing aids&   24 months&   VRChat*, SteamVR\\ 
    \midrule
    
      \textbf{H-P9}&    29/F&   80\% hearing loss &   Hearing aids&   Multiple times&    A social VR app to play games with others, Snapchat*\\ 
    \midrule
    
      \textbf{V-P1}&    28/M&   Blind&  White cane&    Few times&    Horizon Series, Meta Avatars*\\ 
    \midrule
    
      \textbf{V-P2}&    31/F&   Residual vision in one eye &    N/A&   2 months&   Cardboard, Meta Avatars*\\ 
    \midrule
    
      \textbf{V-P3}&    26/M&   Loss of side vision &    N/A &   24 months&   BeanVR, Snapchat* \\ 
    \midrule
    
      \textbf{V-P4}&    38/M&   Blind &   White cane, guide dog &   Multiple times&   VRChat, Rec Room, Oculus Venues, Xbox*\\
    \midrule
    
      \textbf{V-P5}&    32/F&   Only have vision in one eye &   White cane, guide dog&   N/A&    Snapchat*\\
    \midrule
    
      \textbf{V-P6}&    25/F&   Blind&   White cane&   N/A&   Twitter avatar*\\
    \midrule
    
      \textbf{V-P7}&    49/F&   Blind &   White cane&   Multiple times&   A social VR app with educational purpose, several social VR games, Meta Avatars*\\
    \midrule
    
      \textbf{V-P8}&    53/F&   peripheral vision loss &   White cane, guide dog&   N/A&    Meta Avatars*\\
    \midrule
    
      \textbf{M-P1}&    58/F&   Blind since birth; hearing disability since recently &   Hearing aids, white cane, guide dog&   Few months&    A social VR app for education, Meta Avatars*\\ 
    \midrule
    
      \textbf{M-P2}&    38/M&   Blind; has a prosthetic leg &    White cane &   Multiple times&   Rec Room*\\
     \bottomrule
    \end{tabular}
    
    \label{tab:my_label}
\end{table*}

\subsection{Findings}
\subsubsection{\textbf{Experience with Avatars and Social VR Platforms}} 
Sixteen out of 19 participants had social VR experiences. In general, we found that DHH participants had a richer experience with social VR avatars than VI participants. Except for H-P3, all DHH participants had used social VR multiple times. Five DHH participants were frequent users with experiences of more than a year. 
In contrast, VI participants had limited social VR experience, with most of them only trying it a few times. Three VI participants did not have social VR experience, but they used avatars on conventional social media, such as 
Snapchat Bitmoji and Meta Avatars for Instagram. 
Several VI participants mentioned that they attempted to customize their avatars on social VR several times but were blocked when setting up an account. We report the accessibility challenges faced by participants in Section \ref{barriers}. 

The two most commonly used social VR platforms by our participants were VRChat (7 participants) and Rec Room (5). Most participants used social VR via head-mounted VR devices and the most commonly used devices were Oculus Quest, Oculus Rift, and Valve Index. However, three participants (H-P1, V-P2, V-P6) preferred using desktop-based social VR due to the accessibility issues of VR devices.

Participants used social VR for multiple reasons. Most participants used social VR to communicate with friends (7 participants) and play games (6). H-P4 specifically wanted to know other DHH people and connect with DHH community in social VR. Six participants used social platforms for professional purposes, such as hosting VR meetups (H-P1), promoting rights for PWDs as accessibility activists (H-P1),
and using social VR as an education platform (e.g., M-P1, V-P7, H-P4). Notably, three participants (H-P4, H-P6, H-P8) used social VR for American Sign Language (ASL) education. 
They learned or taught ASL in a VRChat community called ``Helping Hands'' \cite{helpinghands}, which was a community for DHH people to communicate via sign language (see details about ASL in VR in Section \ref{sec_VRASL}).

\subsubsection{\textbf{Disability Disclosure via Avatars}} \label{self-presentation} 
Our study indicated that disability disclosure via avatars was an essential strategy of self-presentation for PWD. We identified participants' different disability disclosure preferences in social VR. 

\textbf{Reflect one's physical self.} The majority of our participants (17 out of 19) designed avatars to reflect their physical appearances in real life, including both facial features and outfits. Some participants even hoped to craft the fine details (e.g., makeup, accessories) of the avatars to show their daily styles, habits, and values in real life. 

As part of their physical appearance, eight participants (e.g., H-P3, H-P7, V-P1) expressed their willingness to disclose their disability via their avatars since they believed that the disability ``is part of me'' (H-P3). 
As H-P7 indicated, \textit{``I have [a cochlear implant] on [my avatar] all the time really, just because that's what I do in real life. I like my avatar to represent me as realistic as possible or as close to [myself], so if I have a cochlear implant I'm not ashamed of it.''} Two participants (M-P1, H-P8) specifically emphasized that when they disclosed their disabilities on social platforms, they never aimed to highlight their disability since the disability was just like other physical features, such as hairstyle, skin color, and gender. As M-P1 stated, \textit{``Because that's who I am. I don't see any reason to not disclose [my disability]. It's a part of who I am. And I wouldn't like try to pretend that I was a different race or try to pretend that I was not female cisgender kind of person, so I wouldn't pretend that I didn't have a disability.''} 

Interestingly, although most participants wanted their avatars to reflect their physical selves as much as possible, H-P2 did not want the details to be completely the same: \textit{``I just tried to make it look kind of like me, but not too much like me, because it's kind of creepy when they look like twins. [The avatars] don't ever look completely like you, especially the hair, you cannot replicate someone's hair [with the current VR technology].''} With the limitation of current avatar realism and the Uncanny Valley effect\footnote{Uncanny Valley effect refers to viewer's increased eerie feelings when an entity looks highly human-like \cite{Hepperle2020Uncanny}.}
, she discussed the boundary between the physical self in real life and the virtual image presented by avatars in the virtual world: \textit{``I don't want [avatars] to be too much like me, because it's not real life. Don't try to make it be real life when it's not like [real life].''}

\textbf{Reflect one's physical self with selective disabilities.}
Unlike most participants who preferred disclosing their disabilities entirely, M-P1 selectively disclosed her disability. M-P1 experienced both visual and hearing loss, and she decided which disability to disclose based on the visibility of the disabilities. She used both a white cane and hearing aids in the real life. However, she would only add a white cane to her avatar to signify her visual impairments but not disclose her hearing loss. This is because hearing aids were not visible to people in the real world in most cases. Interestingly, M-P1's consideration of disability visibility was to fulfill the expectation of the audience, specifically people who knew her in real life. As she explained, 

\textit{``I do use hearing aids, but I don't think that they are visible particularly... so I was like, well I am not going to add a hearing aid, because a lot of people probably don't even know that I use [hearing aids], because I don't think they're that visible, whereas people would know that I use a cane. I was trying to make [my avatar] look like me, and I think that if I had hearing aids, a lot of people would be like, why do you have hearing aids, so why did you do that.''}

Moreover, M-P1's experience with different disabilities also influenced her choice of disability disclosure. She preferred to disclose only the ``dominant disability'' that can represent her most: \textit{``I have had the hearing disability for a lot less time than the [blindness], I mean I've been blind all my life, so pretty like used to [blindness]. I think that maybe if I had grown up hard of hearing or deaf that would be more important to represent for me.''}

\textbf{Reflect changes of the physical self.}
Notably, participants who experienced acquired and progressive disabilities mentioned that they wanted to use their avatars to reflect their physical changes and inform people about their current abilities. For example, H-P2, who suddenly experienced one-side hearing loss at the age of 20, emphasized her need to inform people of her current hearing ability via her avatar: \textit{``Because people don't know [my acquired disability], especially when [my hearing loss] happens so suddenly to me. Literally people are like `Oh, you can still hear out with one ear, you're just fine.' But it's like, I have no ability to locate a noise. When there's any sort of background noise, I can't hear anything that you say. The whole world is just like quieter. People don't realize that.''}

\textbf{Present a capable self.}
Uniquely for PWD, some participants wanted to present a capable self via avatars, demonstrating that they were as capable as people without disabilities in accomplishing tasks. However, participants employed divergent strategies to present their capabilities. 

Four participants (H-P4, V-P3, V-P4, M-P2) chose to hide their disabilities in avatars, so that their disabilities would not bias others and overshadow their capability in social interactions. For example, V-P3, who played multiplayer games in BeanVR, preferred not to disclos his visual disability, because he was afraid of being judged as a weak player. 
As V-P3 described, \textit{``Sometimes there are challenges, when you are trying to make friends, people may like disregard your friend request because [your] visual impairment. Sometimes you have to keep it a secret. I was like looking for teammate on BeanVR, so that you can join a game, nobody wanted to be my teammates at the end because maybe my weakness.''} 

Reversely, some participants (V-P1,V-P5) crafted their avatars to show their disabilities and used them as a way to prove their capability and independence in using this new technology. As V-P1 said: \textit{``I see that avatar can be a symbol of hope, a symbol of I can do this. For me, disability is not the end. I like to overcome my disability and show others that I can overcome it. I've thought of ways to overcome every obstacle. So, by [customizing my avatars], [I show that] every person with disability can become independent.''}

\textbf{Present a professional self for disability education and awareness.} 
Nine participants (4 DHH and 4 VI participants, and M-P1)  
disclosed their disabilities to trigger conversations and educate the public about disability and equity. As H-P7 expressed, \textit{``I think if people notice [the disability feature of my avatar], it can spark an interesting discussion about oh look they're wearing a cochlear implant, what is that.''} H-P1 further emphasized the importance of educating people about DHH people: \textit{``Avatars [with disabilities] were kind of cool, because that's me educating people that deafness is a spectrum, we don't offer in buckets.''}

V-P7 shared her disability to advocate for social justice. As V-P7 said, \textit{``Part of the reason [of disclosing my disability] is that, I am an advocate for social justice and all marginalized communities, and so I want people to know that I am an ally, and I am here to support and work for equitable opportunities for all individuals regardless of what their different diversity might be, and so having my disability out there on social media gives people a little more insight into me as a person.''}

\textbf{Disclose disability selectively based on social contexts.} 
Some participants decided whether to disclose their disabilities based on specific social circumstances.
Seven participants (three DHH and four VI participants) considered the audience. Some preferred disclosing their disabilities only to the audience who they knew in real life. For example, H-P1 did not want to disclose her disability to random strangers in social VR: \textit{``I think the only situation I can think of where I may not disclose is when there are other strangers, too many strangers. In a setting where you can just show up, and a stranger can come run after you, I would not want to [disclose my disability].''} 

V-P2 emphasized that disability was her privacy. She would not share information about her disability on every occasion. Instead, she would assess the situation and audience, then make the decision accordingly. As she indicated, \textit{``Well, in general, I am a private person. I don't like too many people knowing about my disability, because they haven't been through what I've been through, they judge before they even know. I am a really sensitive person, so somebody says that to me may not be hurtful to them, [but hurtful to me].''}

\textbf{Present a different self that is not defined by disability.}
While the majority of participants were willing to disclose their disabilities via avatars, three participants (H-P4, V-P4, and M-P2) held the opposite attitudes. H-P4 preferred not disclosing her disability at all because she did not want her disability to become her representative identity and to overshadow her personality when socializing with others: \textit{``I don't like to disclose [my disability], because I don't want that to be the initial impression that people have, that this person is deaf. I just don't feel a need to say anything unless I have to. Because it might just be a personal thing, and I just don't want that to be characteristically associated. It's not a bad thing. But I've been that way (not disclosing disability unless necessary) all my life, not just in VR.''}  

Moreover, some participants (H-P4, V-P4, and M-P2) viewed social VR as a world where they can express themselves freely and explore an ideal self that was different from the real world. As V-P4 said, \textit{``Because I feel like in virtual reality, your disability really shouldn't matter because it's virtual reality. And virtual reality you're not hampered or hindered or shouldn't be any way. So my case of blindness, plus a prosthetic leg, should not hold me back in virtual reality.''}

\subsubsection{\textbf{Avatar Customization for Disability Disclosure.}} \label{tools}
While most participants indicated a strong desire to be able to show their disabilities via avatar design, they experienced various difficulties in practicing disability disclosure via avatars. In this section, we reveal participants' challenges in disability disclosure via avatars, the strategies they adopted, and their desired features to support disability representation. 

\textbf{Challenges in disclosing disability via avatars.}
We identified four main challenges that participants encountered when disclosing disability via avatars. 

First, almost all participants (18 out of 19) complained about the lack of disability representations in avatar design. Seven participants mentioned not knowing or not being able to find any avatar features that represent disabilities on the social platforms they used. While some platforms support limited disability features (e.g., hearing aids and cochlear implants in Meta Avatars), the choices and customization options were limited. 
For example, H-P2 mentioned that Snapchat only provided hearing aids for avatars, but people who used cochlear implants did not want to use them to represent their disabilities. Due to the lack of disability features, V-P5 used regular avatar decorations (i.e., sunglasses) to present her low vision. However, it cannot clearly signify her disability since people without disabilities also use it. As V-P5 said, \textit{``My avatar has sunglasses [since I'm light sensitive], but this doesn't really make sense because anybody can wear sunglasses. I honestly do wish that there is more representation of people with disabilities.''}

V-P4 further emphasized the importance of avatar diversity for disabilities by comparing it to other minority groups, \textit{``I think it's a very good idea to have [disability-related] features, I might even use them. Everyone's talking about diversity and it's all about race and gender. But one common thing that gets left out is disability. I think that you should have the option to represent, whether you want to use that or not it's up to you, but it never hurts to have.''}

Second, current disability signifiers for avatars were not always compatible with other avatar features. 
For example, H-P1 tried to put a cochlear implant on her avatar but ended up not using it since it was blocked by the avatar's hair: \textit{``The [Meta Avatars] do have a cochlear implant you can add. But you can't see it, because my hair kind of covered it. I tried to spare [the hair] in a pony tail to see if you could see the cochlear implants. But you couldn't, so I put it back down.''} 

Moreover, the unrealistic size of the virtual hearing devices also prevented participants from applying them to their avatars. H-P5 explained her experience in Roblox: \textit{``The hearing aids are larger than the character, so if you try to put them on, it's seriously like trying to put Barbie clothes on like a Little Tikes doll. They're just wrong, they're just floating.''}

\textbf{Methods of disclosing disabilities via avatars}.
Participants mostly disclosed their disabilities by accessorizing their avatars with the assistive devices they used in real life. Three DHH participants added cochlear implants or hearing aids to their avatars (H-P1 and H-P7 on Meta Avatars, H-P3 on Snapchat). 
For VI participants, there were no avatar features designed to represent visual impairments on current social VR platforms. Two VI participants (V-P2 and V-P5) thus put dark glasses on their avatars to signify their visual impairments. \textit{``My avatar has sunglasses because when I go out, I always have sunglasses with me, because I'm light sensitive. So that is why my avatar is wearing sunglasses.'' (V-P5).}

\begin{figure} [h]
    \centering
    \includegraphics[width=\linewidth]{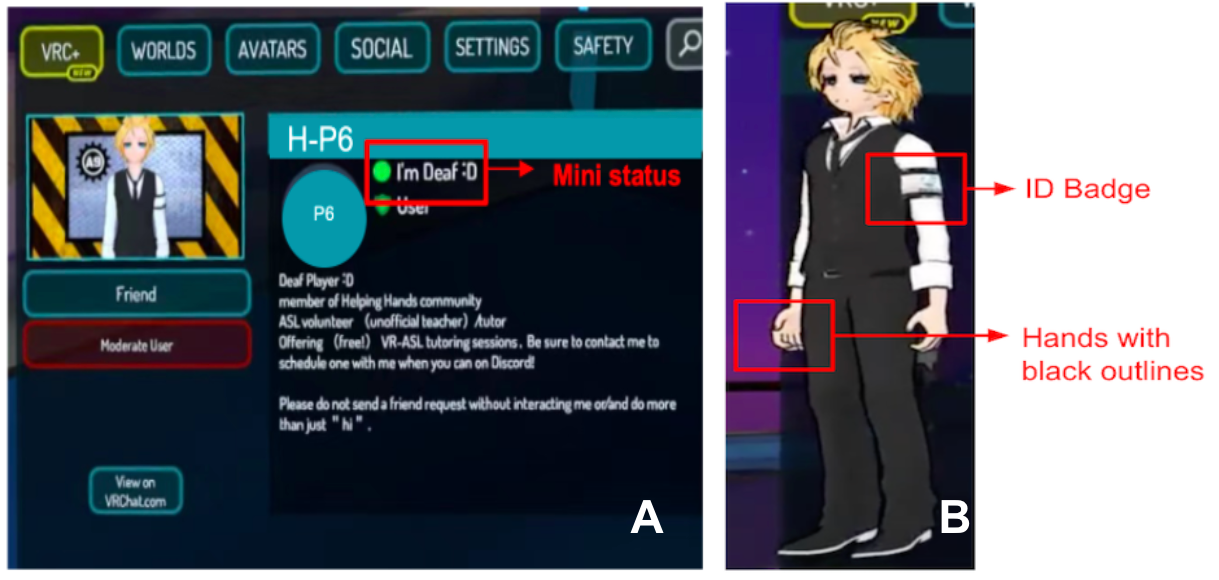}
    \caption{A. H-P6's mini-status in VRChat; B. H-P6's sign avatar with black-outlined hands and ID Badge} 
    \Description{Fig. 2A. showed H-P6’s profile interface in VRChat. The left side of the figure showed a picture of his avatar, and the right side displays his information of user name “H-P6”, mini status of “I’m Deaf :D” with a small circular green icon, the identity of “user”, and profile information which said “Deaf Player :D, member of Helping Hands community, ASL volunteer” and other information about friend request. Fig. 2B. displays H-P6’s full body avatar in VRChat, who is a male figure with yellow hair and pale skin and wearing a black suite. He is standing facing forward 45 degrees to the right. He has an ID Badge on his left arm, which looks like an arm band badge.}
    \label{fig:deaf_avatar}
\end{figure}

Moreover, P-H6 added an ID badge and a mini-status to his avatars to indicate his disability. As he described, \textit{``You can make your own ID badge to explain that you're Deaf or/and Hard of Hearing and have it on your avatar so you can let anyone read it and learn about you. And plus I have a little mini status that would show above my avatar head [if you enabled it in the menu], it'll say `I'm Deaf :D.'''} (Fig. \ref{fig:deaf_avatar}A) 

\textbf{Alternative approaches to disclosing disabilities.} 
Besides avatars, participants (nine VI and seven DHH participants) used other methods, such as bio, posts, photos, emoji, and YouTube videos, to present their disabilities on social media. 


Compared to the 2D social platforms, we found that many participants (six DHH and two VI) were more willing to disclose their disabilities via 3D avatars in social VR. The embodied experience made participants feel more comfortable disclosing their disabilities. In contrast, some participants (H-P4, H-P5) felt that expressing their disabilities in a less embodied way (e.g., via text on social media) was highlighting their disabilities rather than representing themselves. As H-P5 noted, ``The hearing aids are physical reality for me, so I just have them on [my avatar]. I wouldn’t want to put it in text. I wouldn’t want it to be [highlighted]. While [the hearing aid] is visible it feels more subtle [than disclosing my disability in text].'' Meanwhile, some participants felt that the visual presentation of avatars was more noticeable than in other mediums. For example, H-P2 claimed that avatar would be "the first thing" that people see, and H-P7 added that \textit{``it's the easiest way of [disclosing my disability]. I don't think I would put it in like text information like oh I'm deaf. Because I think most of the time well, probably no one reads it.''} 


In addition, H-P7 considered social VR to be a safer place to disclose disability since he could easily move away from toxic conversations. As he indicated, 
\textit{``the freedom of moving around [with avatars] makes you get the power to walk away [from offensive comments], you have a bit of control that you wouldn’t have in social media.''}

However, some VI participants preferred using text-based methods to disclose their disabilities other than avatars since avatars were visual and not accessible. For instance, V-P7 wrote posts or articles to reveal her disability to the public. 
Five VI participants also used emojis (e.g.,    \includegraphics[height=12px]{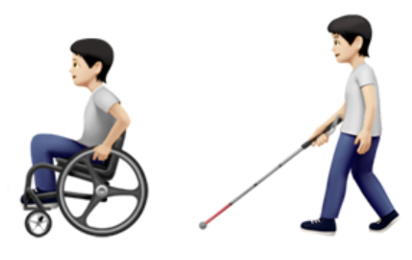}) in their posts to present their disabilities. For instance, M-P1 utilized disability-related emojis in her everyday communications. As she mentioned, \textit{``Well, I didn't create [the emoji], but I like that [they] are already on the iPhone. There is a person with a white cane or whatever. I have used those before to represent me, if I'm writing something, posting a message, or something.''}

\textbf{Desired Disability Representations in Avatar Design.}
Participants suggested various ways to facilitate disability representation in avatars. Most participants hoped that avatar interfaces could include more assistive devices for PWD (e.g., prosthetic limbs, wheelchairs, white cane). While some platforms already provided hearing aids and cochlear implants, DHH participants wanted more options that present different details. For example, they wanted hearing aids with different colors, brands, and wearing methods (e.g., one side and both sides), so that users can freely choose their preferred appearance of assistive devices and feel more realistic. For VI participants, two main desired features were the white cane and guide dog, which were commonly used by VI people in real life. Beyond simply incorporating assistive devices, participants also expressed the need to communicate their personalities through the assistive devices they used. Two participants (V-P1 and V-P6) desired to decorate their avatars' canes. As V-P6 explained, \textit{``[The cane] not only ties in with my disability, but like my person-hood as well around and through that disability.''} Moreover, M-P1 who used Braille wished to have jewelry with Braille on it, which represented ``her culture as a person with a disability.'' 

Besides accessories, H-P1 emphasized the need for a customizable avatar body to accurately reflect PWD's physical appearance, including their disabilities. She used the Helen Keller doll as an example, \textit{``When the Helen Keller doll came out, Haben Girma, who's a very famous blind person, was not happy with that because [the doll's] eyes were proportional. They were the same. But Helen Keller in real life, her eyes are not. I mean, even Haben's eyes are not perfectly proportioned. So maybe that would be something worth improving, is to give people the ability to customize the eyes differently. Some people may want to be as accurate as possible.''}

However, one participant (H-P8) did not support providing disability features for avatars. He was concerned that these features could be abused by people without disabilities and suggested verifying the authenticity of one's disability before a user achieves access to these features. As he described, \textit{``From my experience in VRChat, there are a lot of people who have faked disabilities. Do not trust anyone even your new online friends that you have met recently. I don’t think it’s necessary to show any disabilities on the avatars. People have to show proof that they truly have disabilities. For example, I won’t hold back and sign fast as possible to fake deaf people. They will suddenly show their ugly true colors.''}

\subsubsection{\textbf{Specialized avatars and VR-ASL for the Deaf community.}} \label{sec_VRASL}
\begin{figure*}[h]
    \centering
    \includegraphics[width=\linewidth]{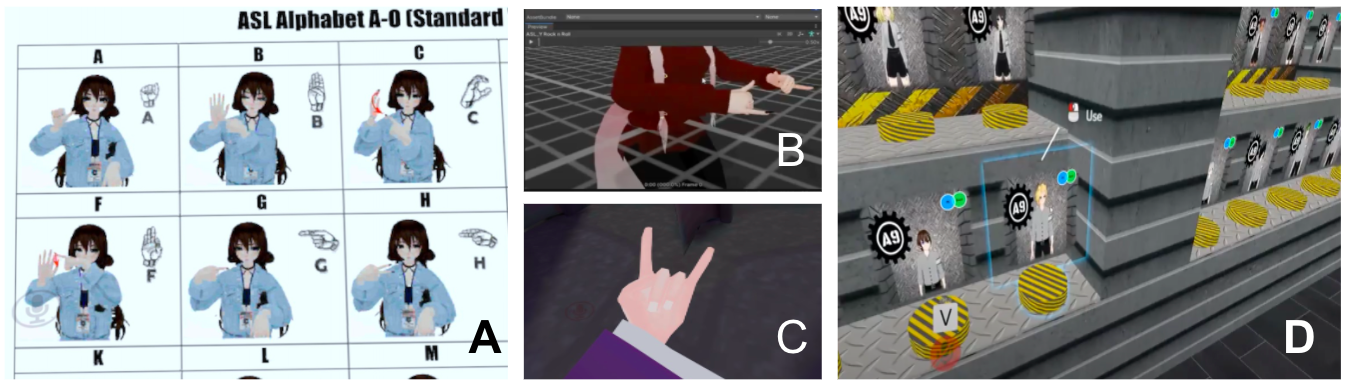}
    \caption{A. Signing the alphabet using ASL vs. VR-ASL; B. H-P4's avatar signing ``y'' hand; C. the ``Rock and Roll'' gesture; D. Warehouse with sign avatars created by ``A9''.
    } 
    \Description{Fig. 3A. illustrates a female avatar signing alphabet “A” in VR-ASL. To the right of the avatar, a hand gesture that shows how to sign “A” in ASL is displayed. These two figures are put in a square to demonstrate the difference between signing “A” in ASL and VR-ASL. This figure contains six such squares to show the difference in signing six alphabets. Fig. 3B. illustrates H-P4’s avatar signs alphabet “Y” in VRChat by holding both hands, palm facing out, with thumb and pinkie finger sticking out and the rest of fingers curled in. Fig. 3C. displays an avatar’s left hand in VRChat, with index finger up, middle fingers down, pinky up, and thumb in, which represents a standard “rock and roll” gesture. Fig. 3D. displays the place that contains many avatars designed especially for signing}
    \label{fig:vr_asl}
\end{figure*}

Notably, three people in the Deaf community\footnote{The Deaf community ``views themselves as a unique cultural and linguistic minority who use sign language as their primary language'' \cite{DeafCommunity}.} (H-P4, H-P6, H-P8) reported using specialized avatars to sign in social VR, specifically in VRChat (Fig. \ref{fig:deaf_avatar}B).  
All three participants were members of the ``Helping Hands'' community in VRChat, where they were either learning or teaching American Sign Language (ASL). Instead of typical ASL, they used VR-ASL, a simplified version of ASL in VR (Fig. \ref{fig:vr_asl}). We report participants' experiences with the specialized avatars and VR-ASL.  

\textbf{VR-ASL.} ASL in real life involved rich movements of fingers, body gestures, and facial expressions. However, not all signs can be recognized in VR due to its limited tracking capabilities. Thus, Deaf users simplified or adjusted some signs so that they could be captured by the current VR controllers, which resulted in VR-ASL (Fig. \ref{fig:vr_asl}A). It is worth noting that all three participants who used VR-ASL adopted the Valve Index Knuckles Controllers, which can capture more hand gestures and finger movements than other VR controllers \cite{knuckles}.

\textbf{Specialized sign avatars.} To enable people to better perceive the signs, participants used specialized sign avatars in VRChat. The sign avatars had three unique characteristics. First, their hands usually had a high contrast skin color to the avatars' clothes, which increased the visibility of the hand gestures.  
As H-P4 mentioned, \textit{``Oftentimes, people who use sign language, if their [avatar] skin is white, it's easier to see and contrast [with dark shirts color], so that helps people visually understand what you're saying.''} Second, many sign avatars had black outlined hands to further enhance their visibility (Fig. \ref{fig:deaf_avatar}B). Last, sign avatars were re-programmed to support richer gestures, so that users can conduct more sophisticated signs in social VR (see the creation of specialized avatars in Section \ref{sec_VRASL}).
For example, H-P4's avatar can perform the hand gesture that signified the letter ``y'' in ASL (Fig. \ref{fig:vr_asl}B), while with the same controller input, typical avatars would only pose a ``rock and roll'' gesture (Fig. \ref{fig:vr_asl}C) 
that did not represent any sign letters.

\begin{figure*}[h]
    \centering
    \includegraphics[width=\linewidth]{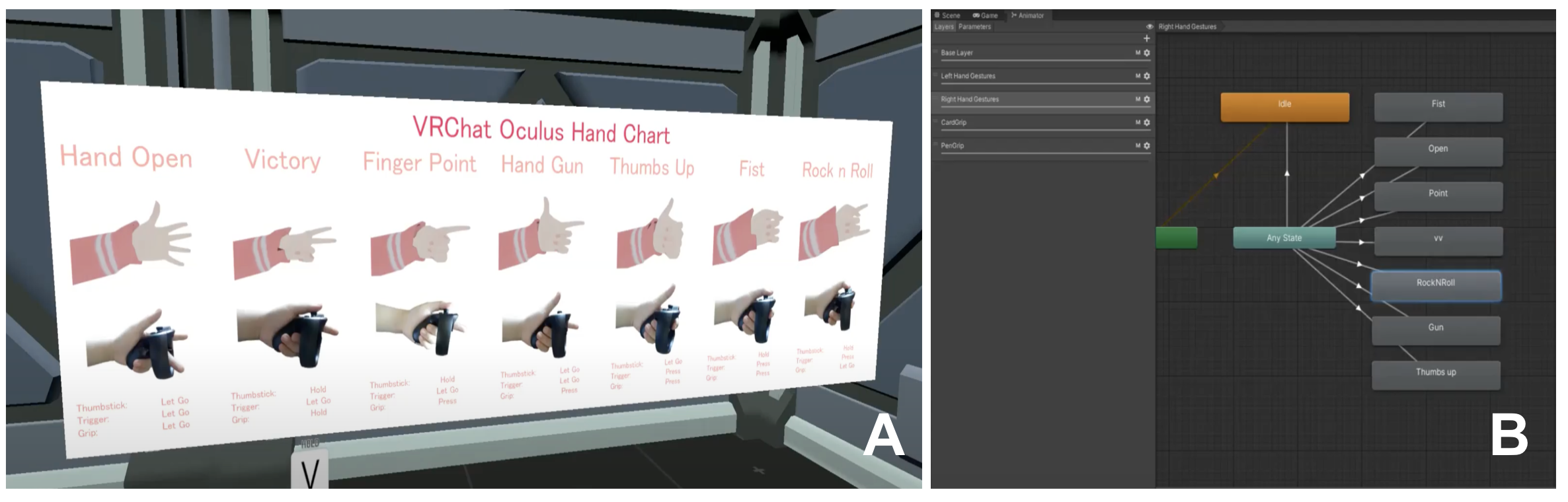}
    \caption{Creation of specialized sign avatars: A. Oculus Hand Chart; B. Unity interface in modifying hand gestures} 
    \label{fig:modify}
    \Description{Fig. 5A. illustrates seven gestures with oculus controllers inputs, with the explanation of how to do the gestures. Fig 5B. Illustrates a tree diagram with gesture layers to customize the oculus controllers inputs}
\end{figure*}



\textbf{Creation of specialized sign avatars.}
Participants used different ways to acquire, create, and customize sign avatars. H-P4 usually imported the publicly available avatars in VRChat into Unity and re-programmed them to support more hand gestures. H-P4 explained the detailed process of her avatar augmentation in Unity: \textit{``I was using custom animations, and I was able to modify the controller inputs to play animations, like I might be able to do `e' hand or flat `o' hand, otherwise you won't be able to have it. Here is a little chart with all the different combinations you can do (Fig. \ref{fig:modify}A), and I was able to modify the game to which I can have access to even more [gestures] with the combination.''} H-P8 also modified the avatars in VRChat, but he asked his friends to do it for him due to technical barriers (see Section \ref{barriers}).

Instead of creating his own avatars, as a teacher who taught VR-ASL in Helping Hands, H-P6 received his avatars as a gift from ``A9,'' a professional creator of sign avatars. A9 owned a warehouse in VRChat, showcasing all the sign avatars he created for VR-ASL (Fig. \ref{fig:vr_asl}D). Members in Helping Hands can choose sign avatars from his warehouse.  

\textbf{Unique disability disclosure via sign avatars.} All three participants who used VR-ASL preferred not using any hearing devices on their avatars, because these assistive devices can not distinguish Deaf people from deaf or hard of hearings people. Instead, the best way to present the Deaf culture was to sign via their avatars. As such, the sign avatars and the use of VR-ASL became a unique way of disability disclosure for Deaf people. As H-P6 explained, \textit{``I can convey and gesture [my disability] to anyone I meet, `Sorry, me deaf, me sign' and they get the gist just fine anyway.''}

\subsubsection{\textbf{Avatar Creation and Accessibility}} \label{barriers}
Avatar creation and customization posed barriers to some participants, especially those with visual impairments. We describe the different challenges DHH and VI participants faced.

\textbf{Barriers to DHH people.} For DHH participants, the avatar customization interface in social VR was accessible and easy to use. Six DHH participants reflected that they did not encounter any difficulties because all avatar features and customization steps were visual and rarely involved any audio information. 

However, the major challenge occurred when creating the specialized sign avatars with third-party platforms since it required programming skills (Section \ref{sec_VRASL}). 
Both H-P4 and H-P8 reported experiencing technical issues and highlighted the lack of support in solving the issues. As H-P4 emphasized, \textit{``Oh so many times, you don't even want to believe, it's always difficult because when the issue is about avatar creation, as of now is that there isn't a lot of resources about it. There isn't a lot of tutorials that are either very relevant or high quality for creating avatars. So that has a lot to be progressed. That's like my biggest issue. I don't know how to do something.''} 

Although some third-party platforms offered tutorials for avatar customization, most tutorials focused on audio instructions and were not accessible to DHH users. As H-P6 complained, \textit{``I think there were some visual examples and others, which was very helpful to learn about, but for most of it was about audio stuff which I, unfortunately, can't hear them. I feel like they haven't thought this out very well.''}

\textbf{Barriers to VI people.} Unlike DHH participants, VI participants faced more significant challenges when creating avatars and using VR platforms in general.
Without the support of a screen reader, VR was not accessible at all to blind users. For example, M-P2 purchased an Oculus Quest but decided to abandon it due to accessibility issues. As he explained, 
\textit{``I have tried, the Oculus Quest, totally inaccessible, because it does not speak. I see it has no screen reading functionality in it. Because I have absolutely no use for this thing unless Meta includes screen reading technology and text-to-speech technology in this. Still, it's sitting in my closet in its original box, it's never even been opened.''}

For low vision participants, similar to Zhao et al.'s results \cite{zhao2019seeingvr}, we found that information readability was a big challenge. The text on some avatar interfaces was too small. Moreover, some interfaces appeared at a fixed distance, preventing users from getting close to see. As V-P4 said, \textit{``For avatar creation, they'll use that sort of flat screen style interface, but then the texts are not large enough, well then you either guess, or you just don't interact with it. So that's where the problem is.''}

\textbf{Avatar creation on social media.} Compared to VR, avatar creation on conventional social media was more accessible to VI participants since participants can access the interface with their phone or computer that incorporated a screen reader. Seven VI participants had experience creating avatars on social media, such as Instagram and SnapChat. However, VI participants still faced various challenges. 

The biggest issue was the lack of sufficient descriptions across the whole avatar design process. When designing avatars, participants (e.g., V-P1, V-P7) had difficulty understanding each avatar feature, tracking their design progress, and confirming the final outcome. Without suitable descriptions and notifications, the avatars created by VI participants (e.g., V-P7, V-P8) did not match their desired avatar appearances. 
As V-P7 indicated, \textit{``Meta has a little bit of description, but not everything has a description. So the first [avatar] I made, [my friends] said it was terrible, didn't look anything like me. I didn't choose the right face, I didn't choose the right color skin tone. People told me the outfit I chose was ugly. I didn't get a lot of positive feedback.''} Since participants could not see and confirm the final look of their avatars, they did not build strong connections with the avatars, thus not caring much about their avatars: \textit{``When I was designing [my avatar], I did have to have my daughter help. It wasn't very accessible and so really it doesn't mean a lot to me because I'm not completely invested in it, because I can't confirm that I like it'' (V-P7).}

Moreover, the avatar customization interfaces usually offered a long list of options, which was difficult for VI users to navigate with screen readers. As V-P6 said, \textit{``I think there were so many buttons to flick through, for example, I would choose my hair color and it will be like 60 of them. Those processes [pose] accessibility hazards. I am good with technology, but I think that someone who was not so good with technology would get really overwhelmed with that.''}


\section{Discussion} \label{discussion}


Our research has contributed the first exploration of avatar diversity and PWD's self-presentation in social VR. We answer the three research questions proposed in the Introduction. 
Firstly, our systematic review in Study I highlighted the lack of disability representation in avatar design on the mainstream social VR platforms. Only the Meta Avatars (and the platforms that adopted the Meta Avatars) provided disability-related features, but the features were restricted to DHH people (RQ1). Secondly, our in-depth interviews in Study II indicated that PWD employed a spectrum of self-presentation strategies via avatars in social VR (see details in Section \ref{discussion: presentation}), and the majority of them showed strong preferences in disclosing their disabilities via avatars by adding the assistive tools they used in real life (RQ2). 
We also uncovered the major differences between DHH and VI participants in avatar perception (Section \ref{discussion:difference}). Lastly, the avatar creation and customization process posed barriers to PWD, especially for VI participants (RQ3). In this section, we discuss the unique self-presentation strategies of PWD from the lens of disability disclosure, the avatar perception differences between DHH and VI users, and the design considerations we derive to inspire more accessible and inclusive avatar design.  

\subsection{Disability Disclosure via Avatars in Social VR} \label{discussion: presentation}
Our research identified some similar self-presentation patterns observed in people without disabilities. We confirmed that most people regarded the embodied avatars as themselves and designed avatars to reflect their physical selves \cite{freeman2021body}, and people may want their avatars to reveal a different self in online games other than their physical selves \cite{kafai2010your}. 

Beyond the insights from prior research, our research uncovered PWD's unique avatar perceptions and self-presentation strategies in social VR. We found that PWD managed their disability disclosure to craft their self-images in social VR. 
Instead of a binary switch between disclosing and not disclosing \cite{davis2021machine}, PWD adopted a spectrum of strategies to determine to what extent and from what aspect they would like to disclose their disability to shape their avatar figures. 
Some participants saw disability as part of their physical selves and wanted to reflect their disability as accurately as possible; some selectively disclosed a certain aspect of their disabilities (e.g., the more visible disability, or the disability with stronger personal attachment) to signify their major disability identity; and some participants with acquired or progressive disabilities employed avatars as a nuanced way to indicate their ability changes. 
Moreover, PWD disclosed their disabilities via avatars to convey certain signals to the audience. For example, some participants designed avatars with disability features to demonstrate their capability and independence, while some participants used their avatars to increase disability awareness and advocate for diversity and equity. 
These patterns demonstrated the importance of disability disclosure for PWD in self-presentation, suggesting the necessity of granting PWD sufficient flexibility to control their disability disclosure in the avatar creation process.

Our research also highlighted PWD's different disability disclosure preferences from other online social platforms (e.g., social media, virtual worlds). While avatars were also supported in some 2D/desktop-based social platforms (e.g., Instagram, Second Life), PWD generally did not feel attached to their avatars due to the lack of embodiment, and many tended to hide their disabilities by using non-human avatars~\cite{stendal2012, davis2021machine}. In contrast, our study suggested that the embodied nature of social VR enabled people to build strong attachment with their avatars, making them more willing to reflect their disabilities in their avatars. Some participants (H-P4, H-P5) mentioned that they did not want to disclose their disabilities on social media since it felt like ``showing off'' their disabilities, but displaying their disabilities via avatars in social VR were more natural. The support of embodied interaction further enhanced PWD's engagement and attachment. Our Deaf participants can thus communicate with their preferred manner in real life---ASL, and they even created VR-ASL to accommodate to the limited capability of current VR technology. Our findings echoed the Embodied Social Presence Theory \cite{mennecke2010} that the embodied avatars and the shared virtual space and activities can affect user perception and bring them to a higher engagement level, and further expanded this theory by providing evidences from the disability perspective. 


\subsection{Avatar Perception of People with Different Disabilities} \label{discussion:difference}
Our findings uncovered DHH and VI people's different avatar experience and perception. Given the visual-driven nature of current VR technology, it is not surprising that VI people face tremendously more challenges than DHH people when designing and using avatars. Our study showed that DHH users had much more substantive avatar experiences than VI users. DHH users already started forming communities in social VR (e.g.,``Helping Hands''), whereas the majority of VI participants only tried social VR for a few times. 
Due to the distinct avatar experiences, DHH and VI people perceived avatars differently. Since VI users cannot see or even imagine the appearances of their avatars, they had weak attachment with their avatars and some ended up not caring about their avatars at all in social VR. As V-P4 said, \textit{``[Avatar design] really is just for fun, I mean it’s just kind of seeing what types of appearances, or what types of features that the avatars have. I really don’t necessarily care about avatars all that much.''} In contrast, DHH people showed stronger attachment with their avatars and spent more effort customizing, specializing, and even re-programming their avatars for better self-presentation and communication. 

While this research focused only on DHH and VI users, our findings indicated that people with different disabilities may have different avatar perception and self-presentation preferences. For example, compared to people who have visible disabilities (e.g., a person with an amputation) or who use apparent assistive technologies (e.g., a VI person who has a cane), people with invisible conditions and using invisible assistive technologies (e.g., a person who experiences chronic pain and uses a trigger tracker) may be more reluctant to disclose their disabilities in social VR. Future research should consider other disabilities and explore different factors that may affect people's disability disclosure decisions, such as visibility of disability and visibility of assistive technology \cite{faucett2017}. 



\subsection{Design Implications for Avatar Diversity and Accessibility}

\subsubsection{\textbf{Avatar Diversity}} We drew three design implications to promote disability representations for avatars.

\textbf{Offer more assistive device representations.} Most participants saw their assistive devices as the key signifier of their disabilities and preferred adding these devices to their avatars for disability representation. 
Participants suggested various commonly-used assistive technologies that can represent their disabilities, including a white cane and guide dog for VI people, cochlear implants and hearing devices for DHH people, and prosthetic limbs, wheelchair, and walking aids for people with motor disabilities. Besides the basic assistive technology representations, participants wanted to further personalize their virtual assistive devices to reflect their aesthetics and personalities, for example, customizing the color and the brand. Adding these features in the mainstream avatar systems would empower PWD to express and present themselves freely and equally in social VR. Designers should involve PWD throughout the whole avatar design process to ensure the suitability of the disability representations.

\textbf{\textit{Support representations for people with invisible disabilities.}} Compared to visible disabilities, some invisible disabilities could not be easily expressed, especially when no apparent assistive technologies are used to signify the disability. In our study, participants suggested using accessories, such as a necklace with Brailles or an arm badge with their community logo, to present their disabilities. We suggest that designers should consider using these indirect mediums, such as outfits with disability signifiers (e.g., a T-Shirt with the Autism Awareness Puzzle logo), to enable people with invisible disabilities to better manage their disability visibility in social VR.

\textbf{Guarantee appropriate use of diversity features.} Adding more avatar diversity features may also pose potential risks. Our participants indicated their concerns on the inappropriate usage of these features by people without disabilities. 
The abuse of diversity features can lead to cyber-bullying and increase the misconception of disability from the public. One participant (H-P8) further suggested conducting strict verification on the authenticity of one's disability before they can access the diversity features.
How to guarantee the proper use of the diversity features is a vital issue that should be considered by the commercial social VR platforms from both the design and the policy perspectives. Researchers and designers should investigate how to set up suitable rules to support interaction freedom without marginalizing and harming vulnerable populations on these new and emerging social platforms. 

\subsubsection{\textbf{Avatar Accessibility}} Our findings highlighted the difficulties faced by VI and DHH people when designing avatars. This echoed the general VR accessibility issues revealed by prior research \cite{zhao2019seeingvr, Franz2021nearmi, Mott2019oppo, Carr2010}, and also expanded the problems from a new avatar creation perspective. We drew three design implications to enhance avatar accessibility. 

\textbf{Combine avatar automation with fine-grained customization.} Avatar creation is not accessible to VI participants due to its heavily visual-driven nature and the complex steps to navigate. Our blind participants usually needed external human assistance in customizing their avatars. Instead of crafting the avatars from scratch, some participants suggested automating the avatar generation process (M-P1, V-P5): the avatar system should automatically generate a baseline avatar based on the user's photo, and allow the user to adjust details based on their preferences. While not new, this feature is supported by very limited number of social VR platforms. In our application review (Study I), only three out of 15 commercial social VR platforms support avatar automation. We suggest that designers should consider enabling multiple avatar creation methods---both auto generation and manual adjustment---to enhance the accessibility of avatar creation and customisation for people with diverse abilities. 

\textbf{Convey avatar design outcomes for VI users.} VI users rely on alternative text to perceive graphic information. Our research highlighted the lack of and the need for alternative text and screen reader support for avatars in social VR. While the avatar design system for conventional social media (e.g., Facebook, SnapChat) could be accessed by VI users via alternative text and embedded screen readers, participants reported challenges in understanding their customization progress and final outcomes, which led to their emotional detachment from the avatars. Artificial intelligence (AI) technology could be considered to recognize the avatars and generate semantic confirmations about the holistic avatar customization outcomes, such as whether the avatar looks like the user.  
However, this could be technically and ethically challenging since categorizing and recognizing human facial features is difficult in both technology \cite{Raji2020concern, Stark2019facial, Dalvi2021Survey} and sociology fields \cite{Dunham2015, hanley2021}. For example, an algorithm may mis-recognize particular avatar or user faces and report inappropriate results. Moreover, how to suitably label and describe human appearance, especially the marginalized groups, is another important question to consider \cite{edwards2021image, hanley2021, bennett2021s}. Besides refining the algorithms and datasets from the computer vision perspective \cite{Taigman_2014_CVPR}, HCI solutions could also be adopted, for example, indicating the potential inaccuracy of the recognized results to users \cite{zhao2018face, macleod2017understanding}, or leveraging human-AI collaboration to achieve more reliable results \cite{guo2018crowd,lee2022opportunities}.

\textbf{Add closed caption for all audio information.} Although the avatar creation and customization process was visual-driven, adding closed caption would still be valuable, especially for DHH users. We believe that with proper caption positioning, closed caption will have an significantly positive impact on the social VR ecosystem.

\section{Limitations and Future Work}
In this paper, we studied the avatar experience and perception of PWD on social VR platforms. We focused on two disability groups---DHH and VI people. Future work should take into account more diverse types of disabilities to get a more comprehensive understanding of PWD's disability disclosure preferences, especially for people with invisible disabilities. 
Moreover, our application review study (Study I) focused on the Oculus Quest 2 platform. Although social VR applications mostly support the same avatar design and accessibility features across VR devices, we acknowledge that different VR devices may offer their own system-level avatars (e.g., Oculus offers the Meta Avatars), which may bring nuances to the review results. Thus, other mainstream VR devices should also be considered to achieve a more comprehensive results.
Finally, future work may complement our qualitative findings with quantitative analysis to investigate what factors may impact PWD's disability disclosure behaviors, such as gender, age, and disability visibility. 



\begin{acks}
We thank the National Federation of the Blind for helping us recruit for our study, as well as the anonymous participants that provided their perspective.
\end{acks}

\bibliographystyle{ACM-Reference-Format}
\bibliography{acmart}

\appendix

\section{Appendix}

\begin{table*}[t]
  \footnotesize
    \centering
    \begin{tabular}{p{2cm}p{2cm}p{2.5cm}p{2cm}p{4.5cm}}
    \toprule
      \textbf{VR Platforms} &  \textbf{Avatar type} & \textbf{Avatar Customization} & \textbf{Avatar Realism}& \textbf{Disability Representation} \\
    \midrule
    
    \textbf{Rec Room}  & 
   \includegraphics[height=9px]{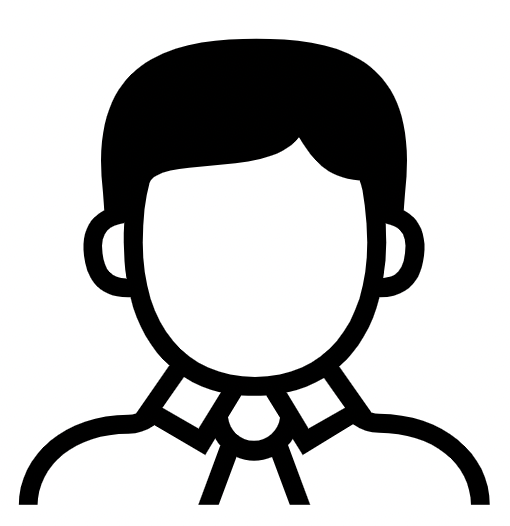}&
   \includegraphics[height=9px]{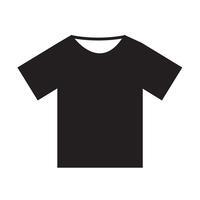}& 
   \includegraphics[height=9px]{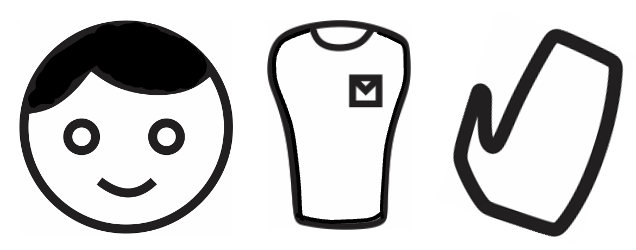} &
   \footnotesize N/A \\ 
    \midrule
    
    \textbf{VRChat} &   
     \includegraphics[height=11px]{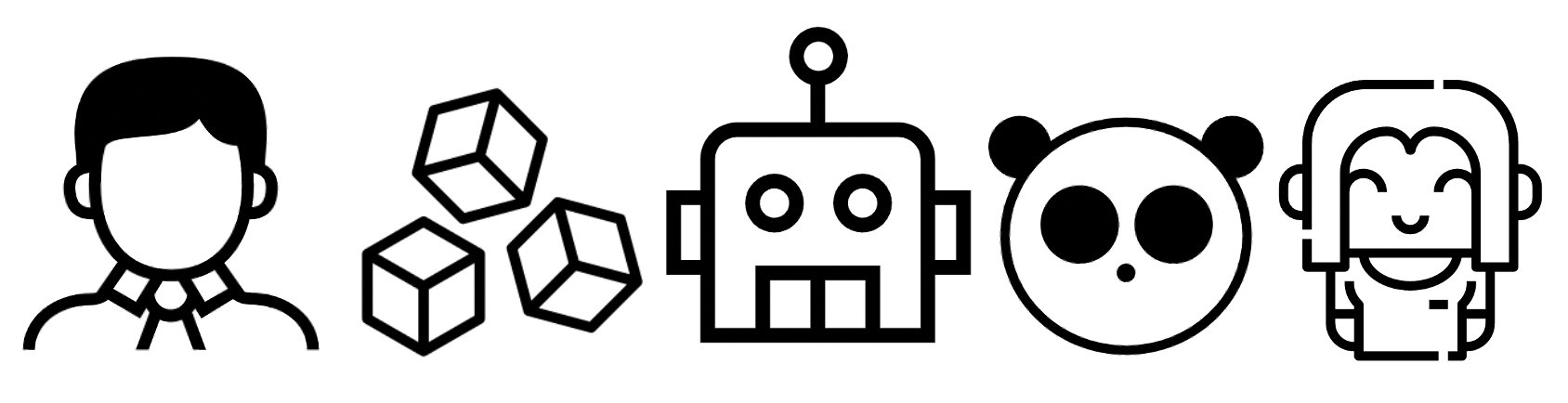}& 
     \includegraphics[height=10px]{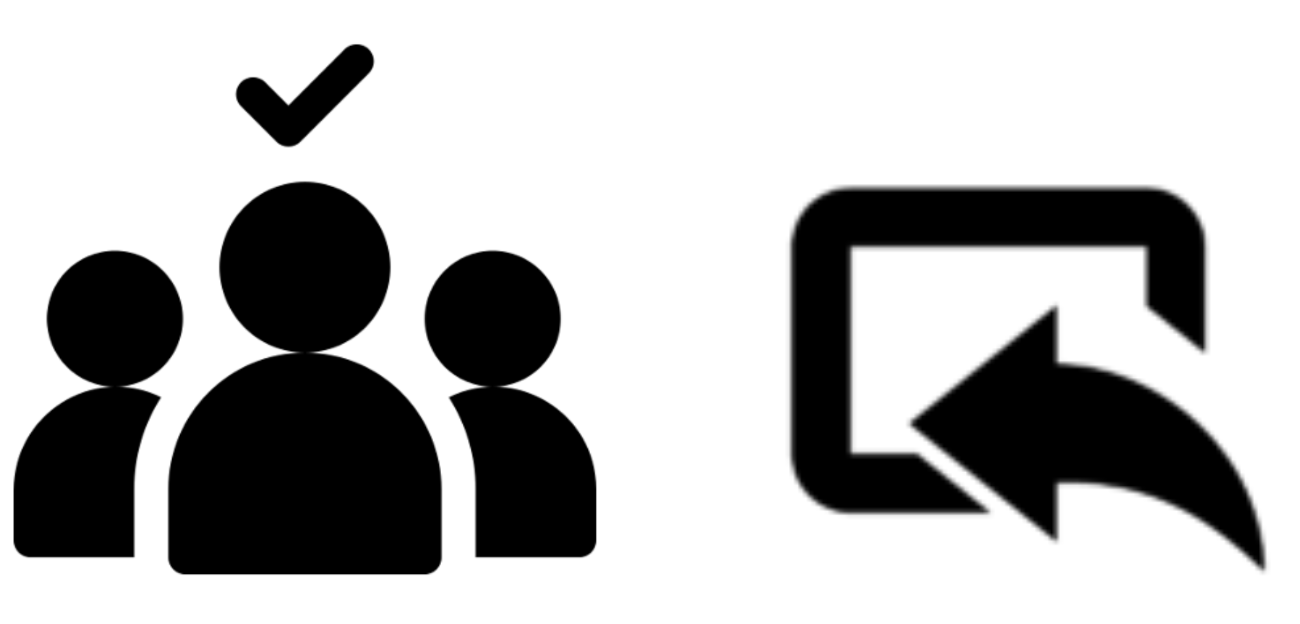}& 
     \includegraphics[height=11px]{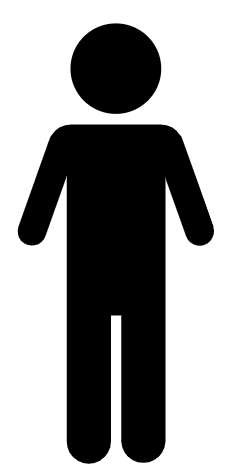}&
    Any uploaded 3rd-party disability features \\   
     \midrule
     
      \footnotesize \textbf{Horizon Worlds} & 
      \includegraphics[height=9px]{sections/images/boy.png}&
      \includegraphics[height=9px]{sections/images/shirt.jpg}& 
       \includegraphics[height=9px]{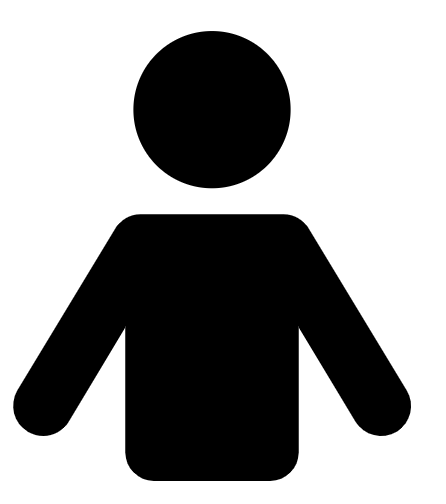}
       &
      Hearing aids; cochlear implants \\
     \midrule
     
     \footnotesize \textbf{vTime XR} & 
     \includegraphics[height=9px]{sections/images/boy.png}&
     \includegraphics[height=9px]{sections/images/shirt.jpg}&
       \includegraphics[height=11px]{sections/images/full.png}& 
    N/A \\
      
      \midrule
     \textbf{AltspaceVR} & 
     \includegraphics[height=9px]{sections/images/boy.png}&
     \includegraphics[height=9px]{sections/images/shirt.jpg}& 
      \includegraphics[height=9px]{sections/images/half.png}&
     N/A \\
     
     \midrule
     \footnotesize \textbf{Bigscreen} & 
      \includegraphics[height=9px]{sections/images/boy.png}&
     \includegraphics[height=9px]{sections/images/shirt.jpg}&
       \includegraphics[height=10px]{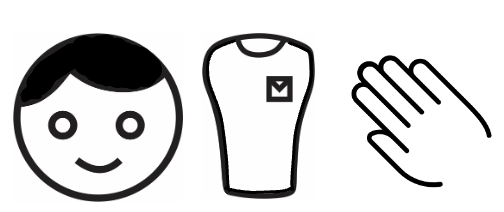}&
      Eye patch \\ 
      \midrule
      
     \textbf{Alcove} & 
    \includegraphics[height=9px]{sections/images/boy.png}&
    \includegraphics[height=9px]{sections/images/shirt.jpg}& 
     \includegraphics[height=9px]{sections/images/half.png}&
     \footnotesize Hearing aids; cochlear implants \\
     
      \midrule
      \textbf{Half+Half} & 
      \includegraphics[height=11px]{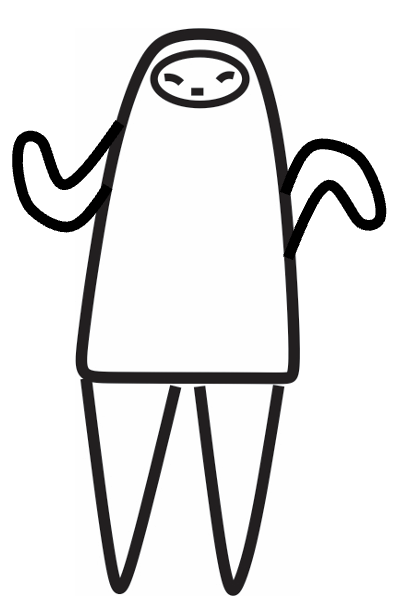}&
      \includegraphics[height=9px]{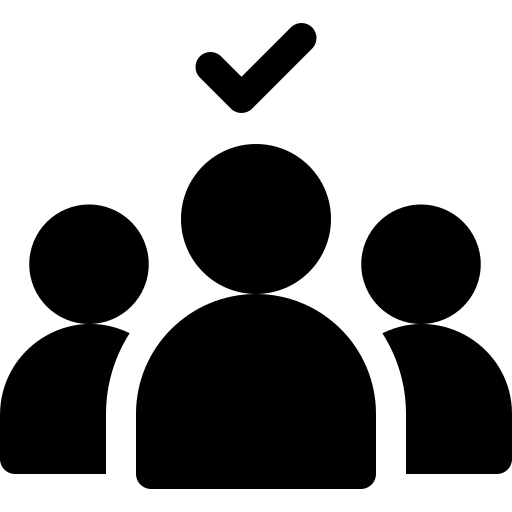} & 
      \includegraphics[height=11px]{sections/images/full.png}&  \footnotesize N/A \\
      \midrule
      
     \textbf{Horizon Venues} & 
     \includegraphics[height=9px]{sections/images/boy.png}&
     \includegraphics[height=9px]{sections/images/shirt.jpg}& 
      \includegraphics[height=9px]{sections/images/half.png}&
      \footnotesize Hearing aids; cochlear implants \\
    \midrule
    
     \textbf{Villa} & 
     \includegraphics[height=9px]{sections/images/boy.png}&
     \includegraphics[height=8px]{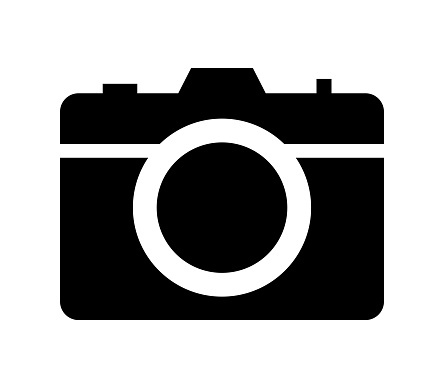}&
      \includegraphics[height=9px]{sections/images/half.png}&
     \footnotesize N/A \\
         \midrule
   
    \textbf{Arthur}  &  
     \includegraphics[height=9px]{sections/images/boy.png}&
    \includegraphics[height=8px]{sections/images/photo.jpg}&
      \includegraphics[height=9px]{sections/images/half.png}&
    \footnotesize N/A \\
        \midrule
        
    \textbf{ENGAGE}  & 
   \includegraphics[height=9px]{sections/images/boy.png}&
   \includegraphics[height=9px]{sections/images/shirt.jpg}& 
    \includegraphics[height=9px]{sections/images/full.png}&
     \footnotesize N/A \\
        \midrule
    
     \textbf{Multiverse}  &  
    \includegraphics[height=9px]{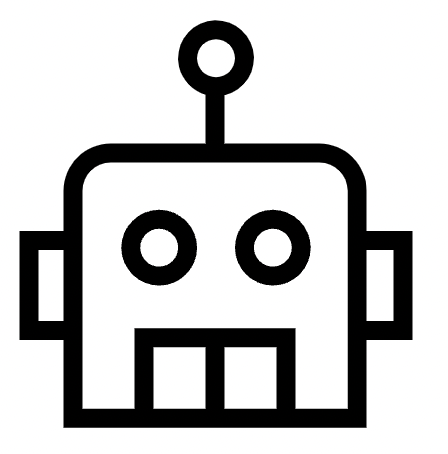}&
    \includegraphics[height=8px]{sections/images/boyy.png} & 
    \includegraphics[height=10px]{sections/images/big_screen.png}& 
    \footnotesize N/A \\ 
        \midrule
    
    \textbf{PokerStars VR}  &  
    \includegraphics[height=9px]{sections/images/boy.png}&
    \includegraphics[height=9px]{sections/images/shirt.jpg}& 
     \includegraphics[height=9px]{sections/images/half.png}&
     \footnotesize Hearing aids; cochlear implants \\
        \midrule
    
    \textbf{Spatial}  &  
    \includegraphics[height=9px]{sections/images/boy.png}&
    \includegraphics[height=8px]{sections/images/photo.jpg}& 
     \includegraphics[height=9px]{sections/images/half.png}&
   \footnotesize N/A \\
        \bottomrule
    
    \end{tabular}
    \caption{Overview of social VR platforms and their avatar creation options. 
    Description of the icons: 
    humanoid avatar= \includegraphics[height=9px]{sections/images/boy.png}, 
    robot avatar=\includegraphics[height=9px]{sections/images/roro.png}, 
    cartoon avatar=\includegraphics[height=9px]{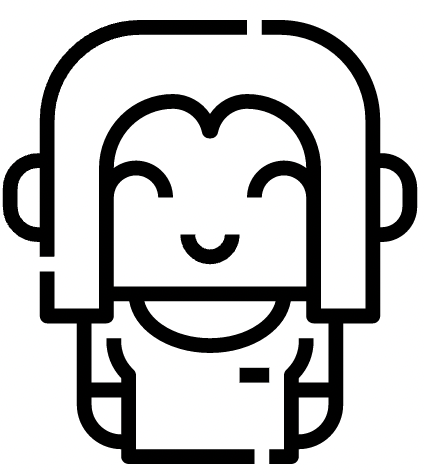}, 
    animal avatar=\includegraphics[height=9px]{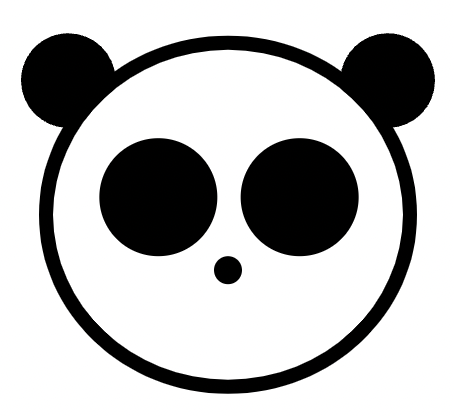}, 
    object avatar= \includegraphics[height=9px]{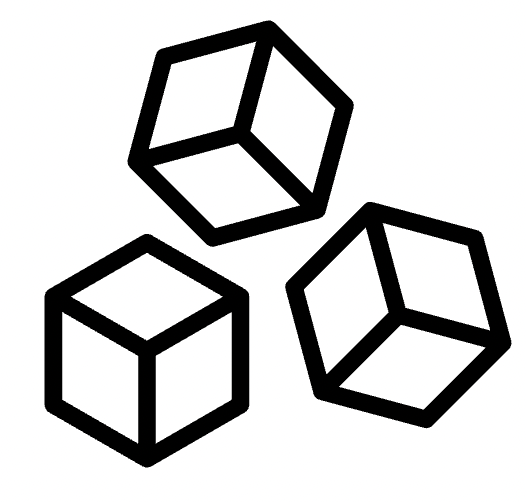}, 
    abstract avatar=\includegraphics[height=9px]{sections/images/half_half.png}, 
    full body avatar includes head, upper body, arm hand, leg=\includegraphics[height=9px]{sections/images/full.png}, 
    partial body avatar includes head, upper body, arm, hand = \includegraphics[height=9px]{sections/images/half.png}, floating head without neck= \includegraphics[height=9px]{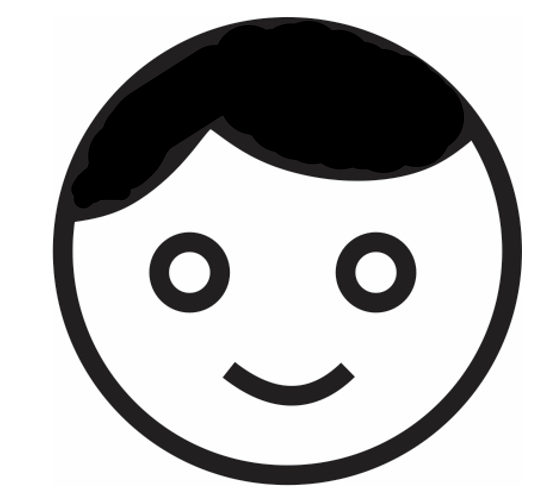}, 
    floating upper body without arm = \includegraphics[height=9px]{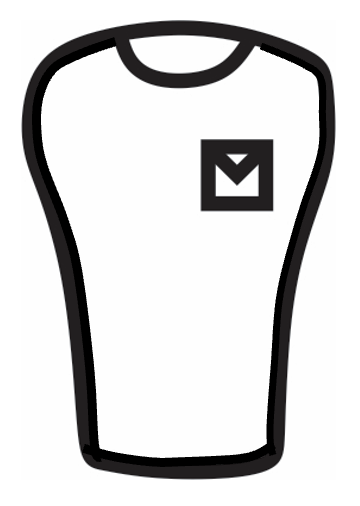},
    floating hand=\includegraphics[height=8px]{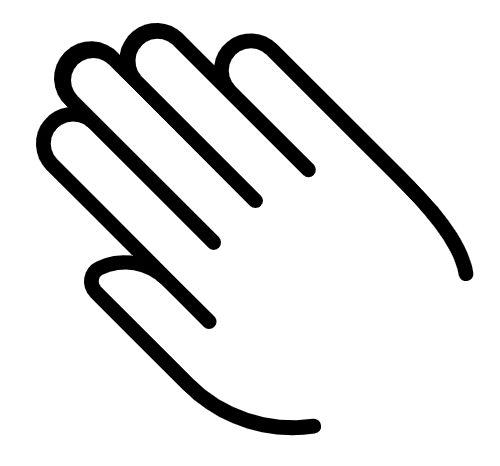}, 
    floating fingerless hand =  \includegraphics[height=9px]{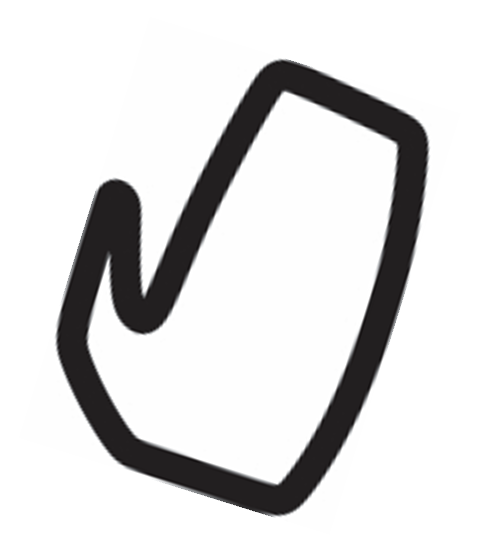}, 
    full avatar selection = \includegraphics[height=7.5px]{sections/images/boyy.png}, 
    avatar feature customization = \includegraphics[height=7px]{sections/images/shirt.jpg},
    photo-based avatar generation = \includegraphics[height=8px]{sections/images/photo.jpg},
    third-party avatar import = \includegraphics[height=9px]{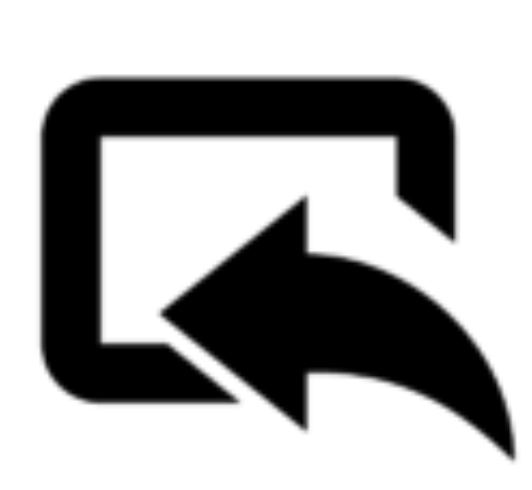} 
    }
    
    \label{tab:socialVRapps}
    \vspace{-5ex}
\end{table*}

\end{document}